\documentclass[aps,prl,twocolumn,floatfix,superscriptaddress]{revtex4-1} 
\usepackage{color}
\usepackage{amssymb}
\usepackage{dcolumn}
\usepackage{amstext}
\usepackage{graphicx}
\usepackage{amsmath}
\usepackage{hyperref}
\usepackage{amsthm}
\usepackage[version=4]{mhchem}
\usepackage{longtable}
\usepackage{amsfonts}
\usepackage{amssymb}
\usepackage{subfigure}
\usepackage{latexsym}
\begin{document}
\title{s$_\pm$ pairing near a Lifshitz transition}
\author{Vivek Mishra}
\address{Joint Institute of Computational Sciences, University of Tennessee, Knoxville, TN-37996, USA.}
\address{Center for Nanophase Materials Sciences, Oak Ridge National Laboratory, Oak Ridge, TN-37831, USA.}
\author{Douglas J. Scalapino}
\address{Department of Physics, University of California, Santa Barbara, CA-93106, USA.}
\author{Thomas A. Maier}
\address{Computer Science and Mathematics Division \&  Center for Nanophase Materials Sciences, Oak Ridge National Laboratory, Oak Ridge, TN-37831, USA.}
\date{\today}
\begin{abstract}
Observation of robust superconductivity in some of the iron based
superconductors in the vicinity of a Lifshitz point where a spin density wave
instability is suppressed as the {hole} band drops below the Fermi energy
raise questions for spin-fluctuation theories. Here we discuss spin-fluctuation pairing
for a bilayer Hubbard model, which goes through such a Lifshitz transition. We find 
s$_\pm$ pairing with a  transition temperature that peaks beyond the Lifshitz
point and a gap function that has essentially the same 
magnitude but opposite sign on the incipient hole band as it does on the electron band that has 
a Fermi surface. 
\end{abstract}
\maketitle

Theories in which pairing in the Fe-based superconductors arises from
the scattering of pairs between hole and electron pockets or between electron
pockets by spin-fluctuations are challenged by the high Tc superconductivity 
reported in mono-layer FeSe films grown on \ce{SrTiO3}.\cite{Wang2012,Se2013,Tan2013,Ge2015}
Scanning tunneling microscopy experiments\cite{Huang2015} as well as ARPES measurements\cite{Tan2013,Zhang2015}
on these FeSe mono-layers find that there are no hole pockets. Furthermore, the ARPES 
measurements of the variation of the gap magnitude around the electron pockets
\cite{Zhang2015} makes the possibility of d-wave pairing, arising from pair scattering 
between the electron pockets, unlikely. However, these experiments also report the existence of
an incipient hole band laying 50 to 100meV below the Fermi energy, implying that the system
is just beyond a Lifshitz transition\cite{LTF1960} where the hole Fermi surface has disappeared. In addition,
photoemission measurements find evidence that superconductivity occurs in the monolayer \ce{FeSe} film, when 
SDW order is suppressed by electron doping \cite{Tan2013} and 
density functional theory calculations\cite{Liu2015} predict that in the absence of electron doping,
the ground state of the mono-layer FeSe film would have SDW order. Thus, it appears that
superconductivity is induced in the FeSe mono-layer when the SDW order is suppressed by
a Lifshitz transition arising from  electron doping or strain\cite{Tan2013}. Motivated by these
results, we have investigated the suppression of SDW order and the
onset of superconductivity near a Lifshitz transition in a two-layer Hubbard model.

\begin{figure*}
\includegraphics[width=0.42\linewidth]{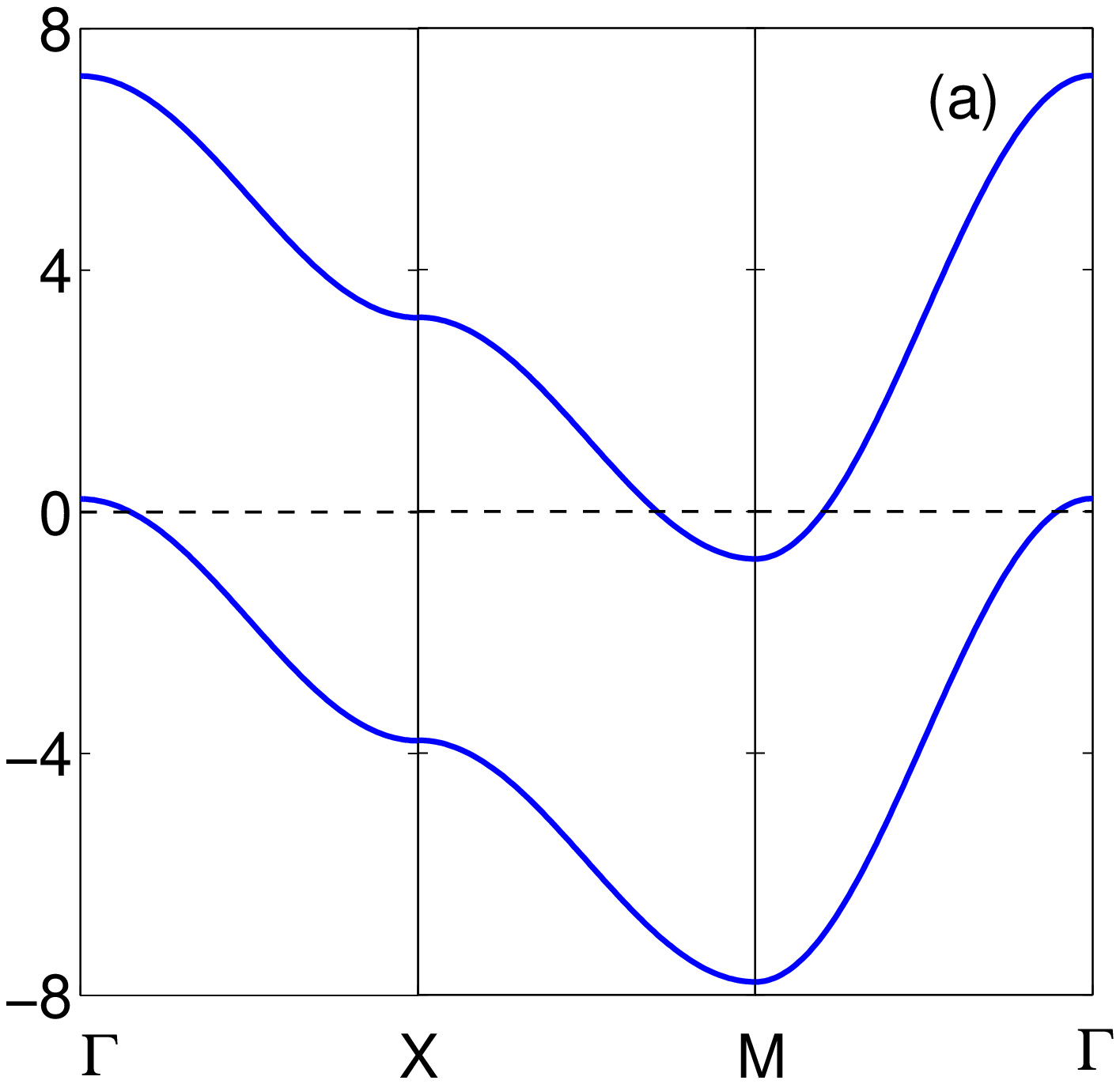}
\includegraphics[width=0.42\linewidth]{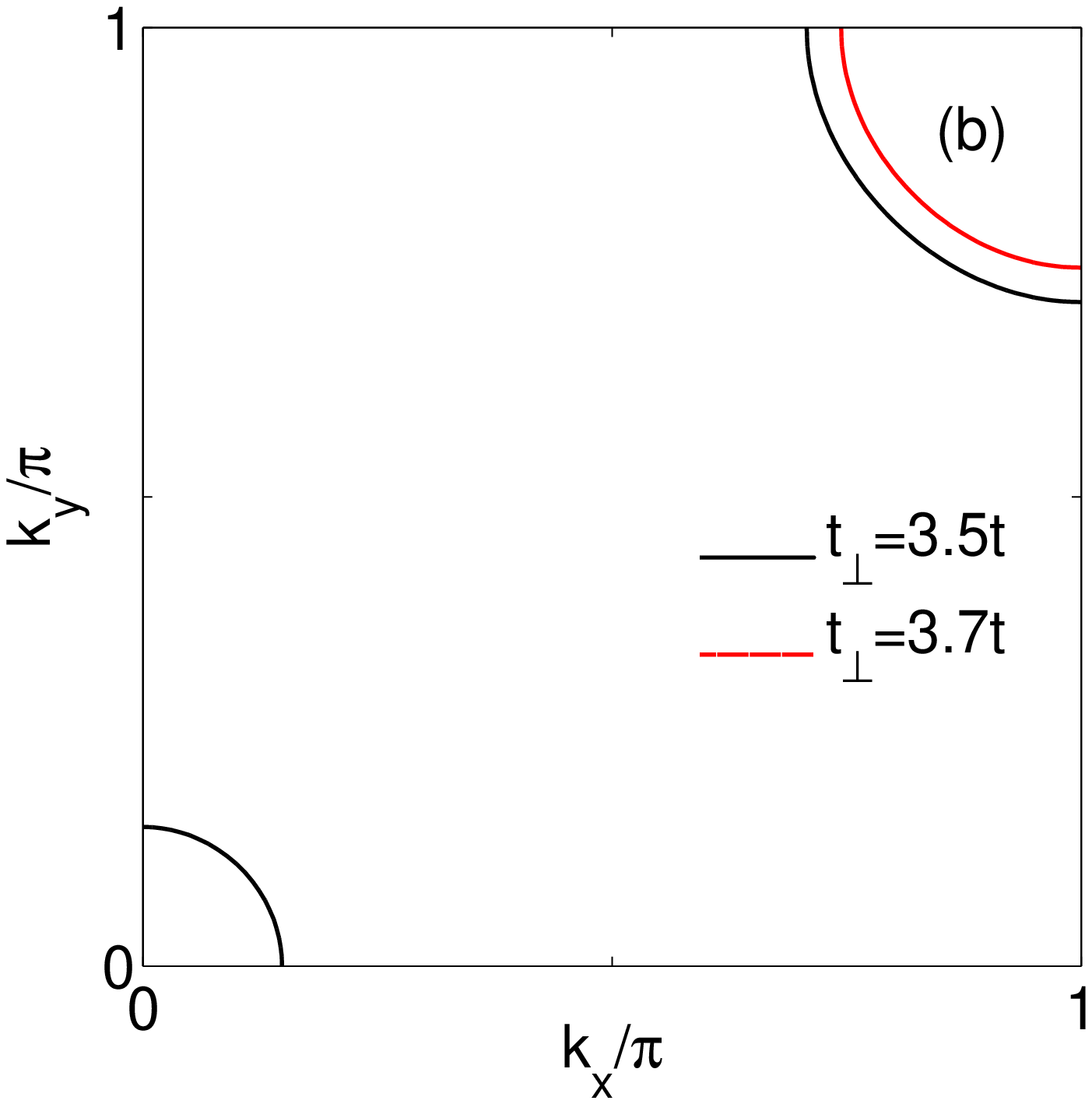}
\caption{(a) The $k_z=0$ and $\pi$ energy bands of the two layer Hubbard model 
plotted ($0,0$) to ($\pi,0$) to ($\pi,pi$) to ($0,0$) with $t_{\perp}=3.5t$ and the chemical potential $\mu$
adjusted for a site filling $\langle n \rangle=1.05$. (b) The Fermi surface for $\langle n \rangle=1.05$ 
with $t_{\perp}=3.5t$ ( solid black) and $t_{\perp}=3.7t$ ( red dashed).}
\label{Fig:FS}
\end{figure*}

The Hamiltonian for the two layer Hubbard model that we study is
\begin{eqnarray}
 \mathbf{H} &=& t \sum_{<i,j>,\sigma,n} \left(c^\dagger_{i\sigma n} c_{j\sigma n} + h.c. \right) - t_{\perp} \sum_{i,\sigma}\left( c^\dagger_{i\sigma 1}c_{i \sigma 2}+ h.c. \right) \nonumber \\
 &+& \mu  \sum_{i,\sigma n} c^\dagger_{i\sigma n} c_{i\sigma n} + U \sum_{i,n}  c^\dagger_{i\sigma n} c_{i\sigma n}  c^\dagger_{i\bar{\sigma} n} c_{i\bar{\sigma} n}. 
\end{eqnarray}
Here $c^\dagger_{i\sigma n}$/$c_{i\sigma n}$ creates/annihilates a fermion with 
spin $\sigma$ in the $n^{th}$ layer (n=1 or 2). The intralayer
hoping is $t$, the interlayer hoping is $t_{\perp}$ and $\mu$ is the chemical potential. The band structure for this model is
\begin{eqnarray}
 \xi_{k} = 2t \left( \cos k_x + \cos k_y \right) - t_\perp \cos k_z -\mu
\end{eqnarray}
with $t_\perp/t=3.5$ and $\mu$ set so that the site filling $\langle n\rangle=1.05$ is shown in Fig. \ref{Fig:FS}(a).
If the filling is kept constant as $t_\perp/t$ is increased, the system has a Lifshitz transition
such that for $t_\perp>3.67$ the hole Fermi surface at the $\Gamma$ point disappears as illustrated
in Fig. \ref{Fig:FS}(b).  We are interested in studying the pairing for parameters such that the spin density wave (SDW)
instability is suppressed by this Lifshitz transition. 

\begin{figure}
\includegraphics[width=.95\linewidth]{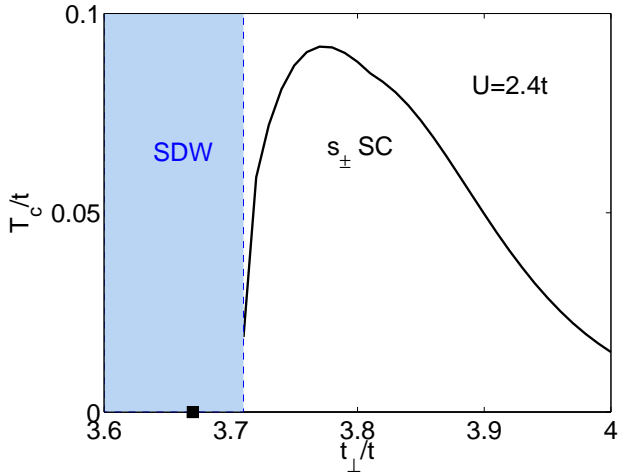}
\caption{The s$\pm$ superconducting transition temperature $T_c$
versus $t_{\perp}/t$ for $\langle n \rangle=1.05$ and $U=2.4t$. The Lifshitz point is denoted by a 
filled square on the $t_{\perp}/t$ axis and the $t_{\perp}/t$ value where the RPA evaluated 
SDW instability ends by a vertical dashed line with light blue shading to the left of it.}
\label{Fig:Tc}
\end{figure}

In a random phase approximation (RPA) the spin susceptibility is given by
\begin{equation}
 \chi^{RPA}(q,\Omega_m)=\frac{\chi_0(q,\Omega_m)}{1-U\chi_0(q,\Omega_m)},
 \label{eq:chirpa}
\end{equation}
with
\begin{equation}
 \chi_0(q,\Omega_m) = \frac{T}{N}\sum_{k,\omega_n}G_0(k,\omega_n)G_0(k+q,\omega_n+\Omega_m).
\end{equation}
Here $T$ is the temperature, $G_0(k,\omega_n)=(i\omega_n-\xi_k)^{-1}$ and $\omega_n=(2n+1)\pi T$ 
and $\Omega_m=2m\pi T$ are the usual fermionic and bosonic Matsubara frequencies.
For a fixed filling, as $t_\perp/t$ is increased and the Lifshitz transition is approached, $\chi_0$
which  peaks near wavevector ($\pi,\pi,\pi$) decreases. 
For $\langle n \rangle=1.05$, we take $U=2.4t$ so that the SDW instability determined from Eq. \eqref{eq:chirpa}
is suppressed by the Lifshitz transition as shown in Fig. \ref{Fig:Tc}. With this suppression of the SDW order,
one can imagine that superconductivity may appear following the usual paradigm. 
However, the Lifshitz transition that has suppressed the SDW instability can also lead to a suppression
of the s$_\pm$ pairing associated with the scattering of pairs between the electron Fermi surface and the 
incipient hole band. For a fixed pairing strength, $T_c$ decreases as the hole band moves below the Fermi energy\cite{ChenPJH}.

To explore this, we solve the Bethe-Salpeter equation
\begin{eqnarray}
 -T \sum_{k^\prime,\omega_m} & &\mathbf{V}(k-k^\prime,\omega_n-\omega_m)G(k^\prime,\omega_m)G(-k^\prime,-\omega_m) \nonumber \\ 
 & & \Phi(k^\prime,\omega_m) = \lambda \Phi(k,\omega_n),
 \label{Eq:Tc}
\end{eqnarray}
and determine $T_c$ from the temperature at which the leading eigenvalue of Eq. \eqref{Eq:Tc} goes to 1.
Here we use a spin-fluctuation mediated interaction,
\begin{equation}
\mathbf{V} (q,\Omega_m) = \frac{3}{2} \chi^{RPA}(q,\Omega_m),
\label{eq:SFP}
\end{equation}
and set $G(k,\omega_n)=\left[i\omega_n-\xi_k-\Sigma(k,\omega_n)\right]^{-1}$ with
\begin{equation}
 \Sigma(k, \omega_n) =T \sum_{k^\prime,\omega_m} \mathbf{V}(k-k^\prime,\omega_n-\omega_m)G_0(k^\prime,i\omega_m).
\label{eq:SE}
 \end{equation}
Note that we keep the Fermi surface unchanged in the dressed Green's function.
For $\langle n \rangle=1.05$ and $U=2.4t$, the resulting
value of  $T_c$, interpolated from the temperature at which $\lambda$ crosses to 1, is plotted in Fig. \ref{Fig:Tc} as a function of $t_\perp/t$.
As shown in this figure, after the SDW instability is suppressed by the Lifshitz transition, a pairing transition
occurs at a $T_c$ which peaks as $t_\perp/t$ increases and then falls off as the hole band is pushed further below the Fermi energy. 

\begin{figure*}
\includegraphics[width=.465\linewidth]{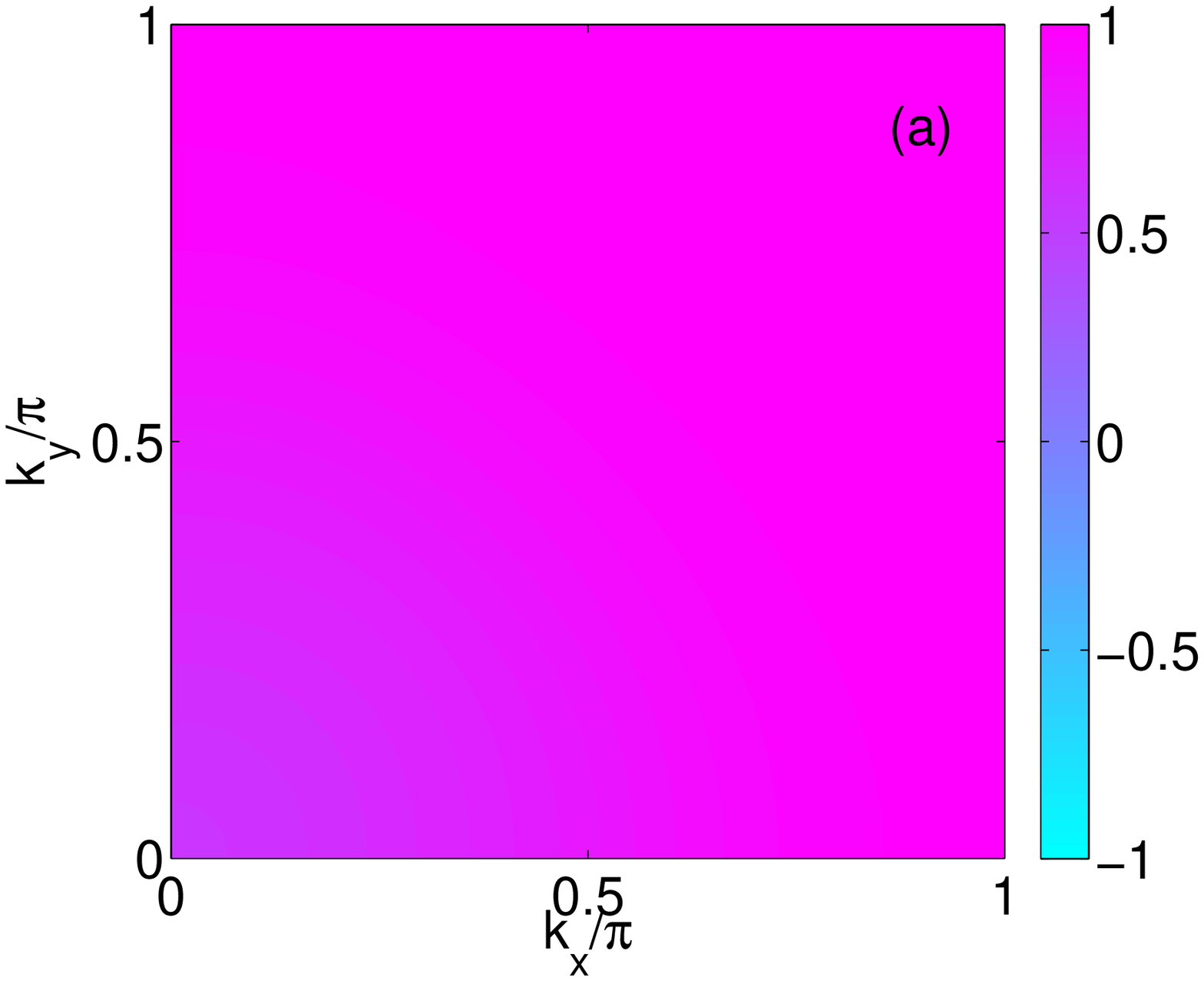}
\includegraphics[width=.465\linewidth]{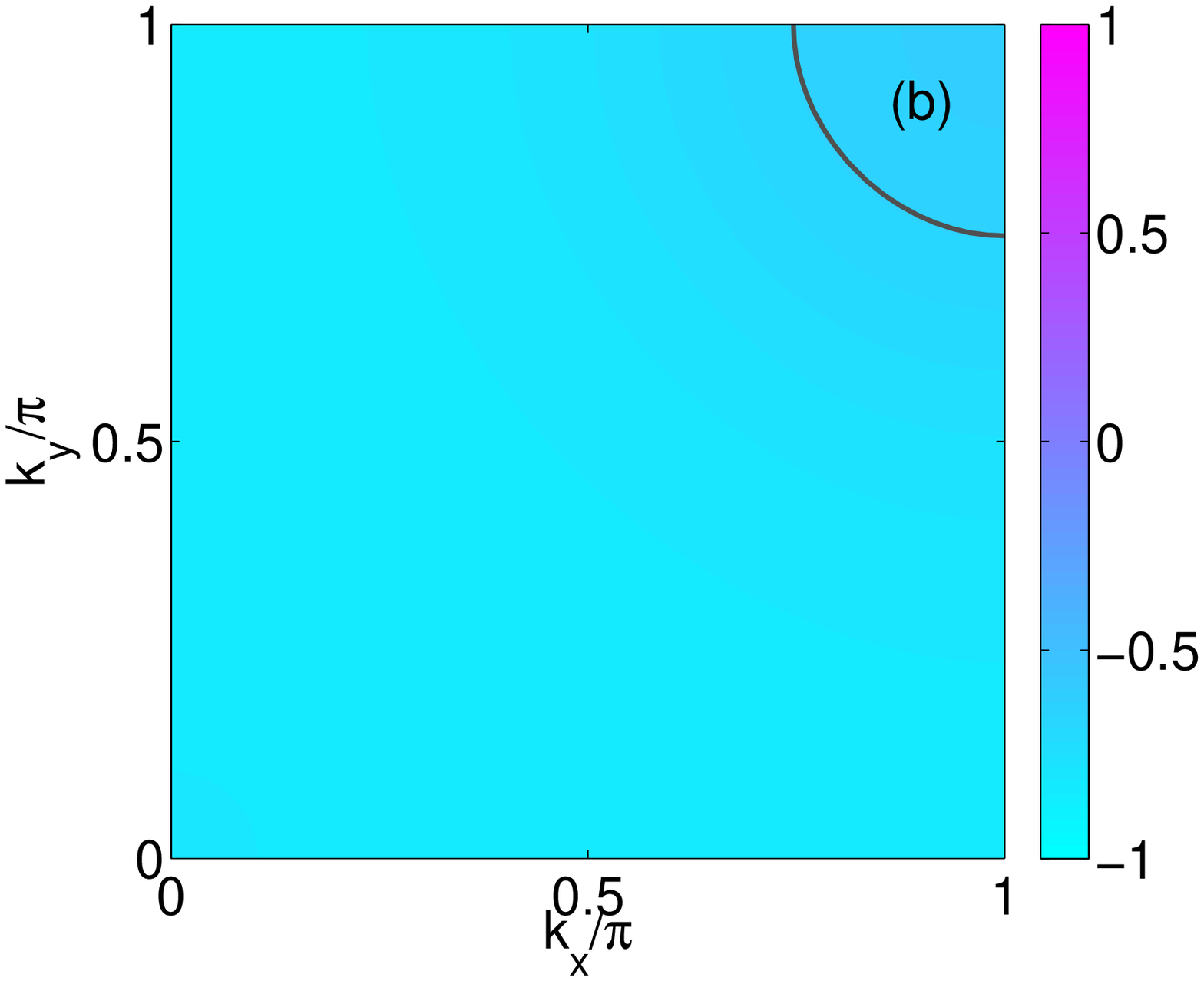}
\caption{The momentum dependence of eigenvectors at the lowest Matsubara
frequency ($\pi T$) for $\langle n \rangle=1.05>$ and $U=2.4t$ at $t_\perp=3.8t$. The eigenvector
is normalized to its maximum value. The momentum dependence for the incipient hole band
is shown in panel (a) and for the electron band is shown in panel (b) with the electronlike Fermi surface.}
\label{Fig:EVEC}
\end{figure*}

The momentum dependence of the superconducting gap function $\Delta(k,\omega=\pi T)\equiv\Phi(k,\pi T)/Z(k,\pi T)$ 
is shown in Fig. \ref{Fig:EVEC}.
This is an $A_{1g}$ (s$_\pm$) state in which the sign of $\Delta$ changes between $k_z=0$ and $k_z=\pi$. One can see that
the magnitudes of the two gaps $\Delta(k_x,k_y,k_z=0)$ and $\Delta(k_x,k_y,k_z=\pi)$ are comparable even though the hole band 
is below the Fermi energy.

\begin{figure}
\includegraphics[width=.965\linewidth]{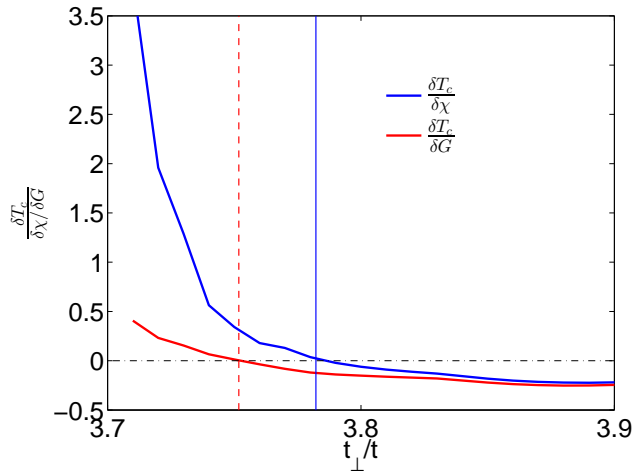}
\caption{The  variation of $T_c$ with changes in $\chi$ (solid blue curve) 
and the pairfield propagators $GG$ ( dashed red curve) versus $t_{\perp}/t$. 
Here one sees that initially as $t_\perp/t$ increases beyond 3.7 and the 
SDW order is suppressed, both the change in $\chi$ and the change in $GG$
lead to an increase in $T_c$. 
Then, as $t_\perp/t$ increases further and the
hole band drops deeper below the Fermi energy, the changes in both $\chi$ and $GG$ lead to a reduction in $T_c$.
}
\label{Fig:FD}
\end{figure}

\begin{figure}
\includegraphics[width=.965\linewidth]{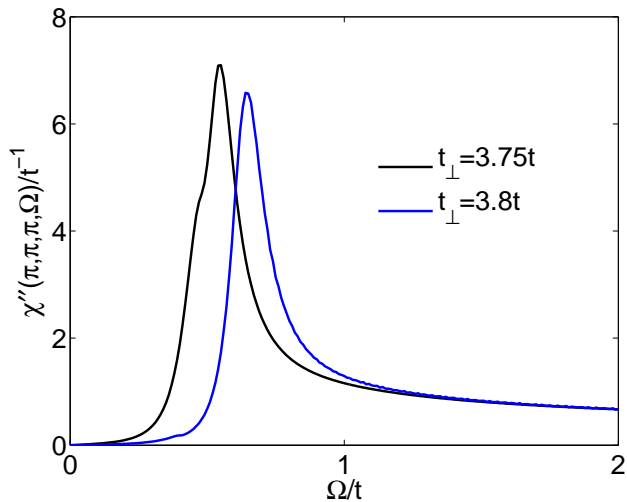}
\caption{The imaginary part of $\chi(\pi,\pi,\pi,\Omega)$ versus $\Omega$ for $U=2.4t$ and $t_{\perp}=3.75t$ and $t_\perp=3.8t$. When the hole
band drops below the Fermi energy, a gap opens in the spin-fluctuation and the spectral weight is shifted to a region that
is more effective in pairing.}
\label{Fig:chi}
\end{figure}

In order to understand the peak in $T_c$ that occurs as the hole band
drops below the Fermi energy, it is useful to examine the dependence of $T_c$ on the 
changes in $\chi(q,\Omega_m)$ and $G(k,\omega_n)$ that occur as $t_\perp/t$ increases
beyond the Lifshitz transition. Viewing the $T_c$ at which the eigenvalue of Eq. \eqref{Eq:Tc}
 goes to 1 as a functional of $\chi(k-k^\prime)$ and the pair propagator $G(k^\prime,\omega_n^\prime)G(-k^\prime,-\omega_n^\prime)$,
  we can calculate the variation in $T_c$ due to the change in $\chi$ when $t_\perp$ increases by $\Delta t_\perp$,
\begin{equation}
\frac{\delta T_c}{\delta \chi} = \frac{T_c\left[ \chi(t_\perp + \Delta t_\perp), G(t_\perp)\right]-T_c\left[ \chi(t_\perp ), G(t_\perp)\right]}{\Delta t_\perp},
\label{eq:FuncChi}
\end{equation}
and the variation due to the change in $G$ when $t_\perp$ increases by $\Delta t_\perp$,
\begin{equation}
\frac{\delta T_c}{\delta G} = \frac{T_c\left[ \chi(t_\perp ), G(t_\perp+ \Delta t_\perp)\right]-T_c\left[ \chi(t_\perp ), G(t_\perp)\right]}{\Delta t_\perp},
\label{eq:FuncG}
\end{equation}
We set $\Delta t_\perp=0.01t$. The results of the calculation are shown in Fig. \ref{Fig:FD}. Here one sees that the initial increase in $T_c$
arises from both the changes in $\chi$ and $G$.  The latter effect is associated with an increase in the quasi-particle
spectral weight $Z^{-1}(k,\omega)$ on the electron Fermi surface that occurs as the hole band drops below the Fermi energy. This increase in the
quasi-particle spectral weight initially ameliorates the decrease in $T_c$ resulting from the submergence of the hole band.
The initial positive contribution associated with the variation in $\chi$ reflects the change in the frequency structure of the spin-fluctuations.
As the hole band drops below the Fermi energy, a gap opens in the low energy $q_z=\pi$ spin fluctuation spectrum
and spectral weight is transfered to higher energies as shown in Fig. \ref{Fig:chi}, which leads to stronger pairing. The ultimate 
decrease in $T_c$ is due to the decrease of the pair propagator $G(k^\prime,\omega_n^\prime)G(-k^\prime,-\omega_n^\prime)$
as  $t_\perp/t$ increases and the hole band drops further below the Fermi energy, as well as the decreasing strength
of the spin-fluctuations.

To conclude, we have studied a  two-layer Hubbard model with parameters chosen so that
a SDW instability is suppressed by a Lifshitz transition in which a hole band at the $\Gamma$ point drops below the Fermi energy. 
Here, we have kept the site filling fixed and varied the interlayer hopping to tune the system through the Lifshitz
point. For a physical system this might by obtained via strain\cite{Tan2013}.
Following the suppression of the SDW order, we find the onset of an s$_\pm$ superconducting state whose transition
temperature $T_c$ initially increases as
the system is pushed beyond the Lifshitz point by further increasing $t_\perp/t$. We find that this increase in $T_c$
is associated with both an increase in the quasi-particle spectral weight and an increase in the strength of the pairing interaction, which are related to
the incipient hole band and the resulting change in the spectral distribution of the spin-fluctuation. We find that the gap function
on the incipient hole band is similar in magnitude but of the opposite sign to that on the electron band which
crosses the Fermi surface. We show that this gives rise to a shift in the neutron scattering spectrum in the superconducting
state, which is related to the well known spin-resonance peak for the usual s$_\pm$ state.

\begin{acknowledgments}
Research sponsored by the Laboratory Directed
Research and Development Program of Oak Ridge National Laboratory, managed by
UT-Battelle, LLC, for the U. S. Department of Energy. 
DJS and TAM acknowledge the support of the Center for Nanophase Materials Sciences, a US DOE Office of Science User Facility.
We acknowledge the Valinor cluster for computational resources. 
We thank  A. Linscheid, S. Maiti, P. Hirschfeld for useful discussion.
\end{acknowledgments}

\end{document}